\documentclass[a4paper]{article}
\usepackage{chicago}
\usepackage{graphicx}
\usepackage{geometry}
\newcommand{\be}{\begin{equation}}
\newcommand{\ee}{\end{equation}}
\newcommand{\bd}{\begin{displaymath}}
\newcommand{\ed}{\end{displaymath}}
\newcommand{\bea}{\begin{eqnarray}}
\newcommand{\eea}{\end{eqnarray}}
\newcommand{\bi}{\begin{description}}
\newcommand{\ei}{\end{description}}
\newcommand{\bq}{\begin{quote}}
\newcommand{\eq}{\end{quote}}

\def\fo{\footnote}

\def\b{\beta}

\def\C{\Chi}

\def\G{\Gamma}

\def\e{\epsilon}
\def\ae{\eta}

\def\th{\vartheta}

\def\m{\mu}

\def\Om{\Omega}
\def\l{\lambda}

\def\p{\varphi}

\def\ph{\phi}
\def\pa{\partial}
\def\Z{{\sf Z\kern-4.5pt Z}} 
\def\R{{\sf R\kern-5.0pt I}}
\def\Q{{\sf C\kern-5.0pt Q}}
\def\C{{\sf C\kern-5.0pt C}}

\def\curl{{\rm curl\,}}

\def\csch{{\rm csch\,}}
\begin{document}
\bibliographystyle{chicago}
\twocolumn[

\author{Alexander~Unzicker\\
         {\small Pestalozzi-Gymnasium}\\
            {\small Melssheimerstr. 11, D-81247 M\"unchen, Germany}\\
            {\small aunzicker@lrz.uni-muenchen.de } \\ [4mm]
Karl Fabian\\
         {\small Department of Geosciences,
        University of Bremen}\\
           {\small Box 330440, D-28334 Bremen, Germany}\\
    {\small	kfabian@uni-bremen.de }}

\title{Displacement Field and Elastic Energy
of a Circular Twist Disclination
for Large Deformations - an Example how to Treat
Nonlinear Boundary Value Problems
with Computer Algebra Systems}

\maketitle

\begin{abstract}

A circular twist disclination is a nontrivial example of a defect in an elastic
continuum that causes large deformations.
The minimal potential
energy and the corresponding displacement field
 is calculated by solving the
Euler-Lagrange-equations. The nonlinear
 incompressibility constraint is rigorously taken
into account.

By using an appropriate
curvilinear coordinate system
a finer resolution in the regions of large deformations is obtained
and the  dimension of the arising nonlinear
PDE's is reduced to two.
The extensive algebraic calculations that arise are done by a
computer algebra system (CAS).
The PDE's are then solved by a difference scheme
using the Newton-Raphson algorithm of successive
approximations for multidimensional equations.
Additional features for global convergence are
implemented.
To obtain basic states that are sufficiently close
to the solution, a one dimensional linearized version of the equation is solved
 with a numerical computation
 that reproduces the analytical results of \citeN{Mur:70}.

With this method, rigorous solutions of the nonlinear
equations without any additional simplifications can be obtained.
The numerical results show a contraction
of the singularity line which corresponds to the well-known
Poynting effect in nonlinear elasticity.
This combination of analytical and numerical
computations proves to be a versatile method
to solve nonlinear boundary value
problems in complicated geometries.

\end{abstract}

\vspace{1.0cm}]


\section{Introduction}
Topological defects in elastic media
usually generate large deformations which require  finite elasticity for their description.
The theory of nonlinear elasticity goes back to \citeN{Cau:827},
and significant contributions came from \citeN{Lov:27}, \citeN{Sig:30},
and for the incompressible case, from \citeN{Riv}. For an
introduction to nonlinear elasticity, the reader may consider
\citeN{Bea:87} or the standard textbooks \citeN{Tru:60} and
\citeN{Tru:65}.

Topological defects, namely dislocations and disclinations,
have been
investigated by means of continuum theories
(\citeNP{Kro:59}; \citeNP{Kro:60}) that were developed after
 the discovery of \citeN{Kon:52}  and \citeN{Bil:55}
that dislocation density is related to the differential geometric
concept of torsion.
After the  result of  \citeN{Fra:49}\footnote{He showed that the energy
increases relativistically with the dislocation velocity.} the
elastic energy of topological defects in elastic
continua has been considered in various publications
on dislocations and disclinations (\citeNP{Esh:49};
\citeNP{Kru:60};  \citeNP{Mur:70}; \citeNP{Nab:79}).
\citeN{Pun:96} discussed
topological defects in a field theoretic context. 

However, calculations of the elastic energy
of topological defects
 have always  been restricted to the linearized equations.
In this case the problem can be  treated
using tensor analysis and stress function tensors
of first and second degree \cite{Kro:59}.
Two  assumptions
usually enter these approximations:

Firstly, one restricts to linear elasticity in the sense
that shearing stresses are assumed to cause a state of simple
shear in the material (as in  \citeNP{Mur:70}).

 Secondly, the energy is assumed to
be a function of the distorsion tensor that can be
approximated by the gradient of the displacement field.
Then, by applying Stoke's theorem, an integration over the
boundaries only can be performed.
Regarding the first point, however, it can be shown
 (e.g. \citeNP{Riv:48a}, p.~467),
 that in the general case
shearing stresses alone cannot maintain a state of simple shear
in the material. Also the second assumption is
justified for {\em small\/} deformations only (\citeNP{Riv:48a},
 p.~476).
In the region close to the defect core, where the
deformations are large the linear
approximation fails to predict finite energies. Thus, these results
are reasonable only outside the core region in where the energy density
diverges.

Large deformations in crystals have been
investigated rarely \cite{Fra:51},
and if, not by means of analytical but by general topological methods
\cite{Rog:76} without considering the elastic energy.

There are several proposals how to treat nonlinear
effects in elastic media with defects. \citeN{Teo:82}
obtained second-order effects for a straight screw dislocation
by applying Willis' (\citeyearNP{Wil:67}; 
\citeyearNP{Wil:70})
scheme which goes back to \citeN{Sig:30}.
This requires, however, certain physical assumptions for every
additional order of approximation.
\citeN{Guo} modelled numerically the large deformations
of a hyperelastic membrane. The cylindrical symmetry
allowed a the reduction to a onedimensional PDE.

Lazar (\citeyearNP{Laz:02b}; \citeyearNP{Laz:02a}) 
obtained solutions for edge and screw dislocations in an
elastoplastic theory.

Povstenko (\citeyearNP{Pov:95}; \citeyearNP{Pov:00}) 
has treated the twist disclination with a nonlocal modulus.
Even if his method involves numerical integration,
it is based on an analytical derivation of the stresses,
and does not allow a contraction of the singularity line.

Here a more  general approach is proposed:
The total elastic energy, i.e. the integral over the energy density
must be a minimum under variation of the displacement field ${\bf u}$.
Additional constraints - like in the present case
incompressibility - are included by
means of a Lagrange multiplier.
If the energy depends on  first derivatives of ${\bf u}$ only,
this leads to Euler-Lagrange equations of second order,
even if the method given here allows the treatment
of higher order equations.

The general method outlined above
is applied to a circular
twist disclination \cite{Mur:70} in an incompressible
hyperelastic continuum.
It is a numerically automatized
process for obtaining rigorous solutions of the nonlinear
equations without special physical assumptions.

Section~\ref{ana} gives a brief introduction to nonlinear
continuum mechanics, followed by a description of the
circular twist disclination. The appropriate
coordinate system and the respective transformation of the
the displacement vector is also explained there.
 All the numerical issues are addressed
in section~\ref{num}, and the results are discussed
in section~\ref{res}.

\section{Analytical description}
\label{ana}

\subsection{Basic concepts of nonlinear continuum mechanics}

Nonlinear elasticity was founded in \citeyearNP{Cau:827} by Cauchy.

One assumes an undeformed, euclidic
continuum with Cartesian coordinates ${\bf  X}= (X, Y, Z)$
(the so-called `reference configuration') and attaches
in every point a displacement vector ${\bf u}$ that points
to the coordinates ${\bf x} = (x, y, z)$ of the deformed
state (`configuration'): ${\bf u} = {\bf x} - {\bf X}$.

From ${\bf u}$ one deduces a quantity
 that transforms the coordinates ${\bf X}$
of the undeformed state to those ${\bf x}$ of the
deformed state: the deformation gradient
\bd
{\bf F} := \frac{\partial {\bf x}}{\partial {\bf X }},
\ed
or, in components,
\be
{\bf F}: = \left( \begin{array}{ccc}
1+u_X &  u_Y & u_Z \\
v_X & 1+v_Y & v_Z \\
w_X & v_Y &  1+w_Z\\
\end{array} \right),
\label{F}
\ee
where $(u,v,w)$ denote the components of ${\bf u}$ and
the subscripts differentiation.
The symmetrical tensor
\be
{\bf B}: = {\bf F F^T},
\label{cauchy}
\ee
is called (right) Cauchy-Green tensor.
It is convenient to introduce the so-called
{\em principal invariants\/} $(I_1, I_2, I_3$)
of a tensor that are defined as follows:
\bea
\label{i1} I_1 = \l_1 + \l_2 + \l_3 \qquad ({\rm trace})  \\
\label{i2} I_2 = \l_1 \l_2 + \l_2 \l_3 + \l_3 \l_1 \\
\label{i3} I_3 = \l_1 \l_2 \l_3 \qquad ({\rm det})
\eea
(the $\l_i$ are the eigenvalues of ${\bf B}$).
Then, in general, the energy density $W$ is a
function of the principal invariants
\be
W = W (I_1, I_2, I_3)
\label{en}
\ee
 of the Cauchy tensor ${\bf B}$ (e.g. \citeNP{Bea:87}), and the
 stress tensor ${\bf T}$
(defined as traction per surface element)
 is given by the constitutive equation
\be
{\bf T} = \b_1 {\bf B}+ \b_0 {\bf E} + \b_{-1}  {\bf B^{-1}}
\label{stress}
\ee
with the $\b_i (I_1, I_2, I_3)$
being {\em functions\/} of
the principal invariants. For small
deformations, i.e.
for $I_1$'s close to 1, and $I_2. I_3$ close to 3, the
$\b$'s have a fixed values - that can be
related to the known elastic
{\em constants\/} in linear elasticity.

\subsection{Incompressibility}

The nonlinear condition of incompressibility is given by \be I_3 =
\l_1 \l_2 \l_3 = det \ {\bf B} = (det \ {\bf F})^2 =1.
\label{incomp} \label{ndiv} \ee and {\em not\/}, as many texts on
linear elasticity state, $div \ {\bf u = 0}$. As a consequence,
the elastic energy $W$ depends on the principal invariants $I_1$
and $I_2$ of  ${\bf B}$ only and eqn.~\ref{en} reduces to \be W =
W (I_1, I_2) \label{eninc} \ee In the present paper,  the special
case \be W(I_1) = C (I_1 - 3) \label{MR} \ee is considered, which
is called a {\em Neo-Hookean\/} material\fo{This corresponds to
the condition $\b_i = const.$ 
 Note  that in finite
elasticity the energy density depends upon the functions $\b_i$.
Therefore,  no general `canonical' energy density exists.} where
$C$ is a constant.

\subsection{Strain-Energy function expressed by displacements}

Considering again a Cartesian coordinate system
and taking into account (\ref{F}), (\ref{i1}) and
(\ref{MR}), the energy per unit volume for an incompressible,
neo-Hookean material reads
\be
W = C (\e_{xx}+\e_{yy}+\e_{zz}),
\label{egy}
\ee
where $C$ corresponds to the shear
modulus $\m$,\fo{In
an incompressible material, $3 C$ equals Young's modulus $E$.}
and the respective components of strain are given by (\citeNP{Cau:827};
\citeNP{Lov:27}, p.~60 and \citeNP{Riv:48}, p.~461)
\bea
\e_{xx}=u_x+\frac{1}{2} (u_x^2+v_x^2+w_x^2)\\
\e_{yy}=v_y+\frac{1}{2} (u_y^2+y_x^2+w_y^2)\\
\e_{zz}=w_z+\frac{1}{2} (u_z^2+z_x^2+w_z^2).
\eea
Hereby ${\bf u}=(u,v,w)$ denotes the displacement vector in $x$-, $y$-, and
$z$-directions and $u_x$, $u_y$ ... etc. the respective partial
derivatives.

\subsection{Euler-Lagrange equations}

The energy of the topological defect can be found by minimizing $W$ under the
variation of the
displacement field $ {\bf u}$ which fulfill the correct spacial boundary conditions.
In addition, the nonlinear constraint (\ref{ndiv}) is taken into account
by means of a Lagrange multiplier $\l$. This requires the minimization of the
Lagrangian
\be
L = W + \l
(det \ {\bf F} - 1). \label{LL}
\ee

 The Euler-Lagrange
equations are thus obtained from \bea \label{el1} \frac{\pa L}{\pa
u} -\frac{d}{d x} \frac{\pa L}{\pa u_x}
-\frac{d}{d y} \frac{\pa L}{\pa u_y}-\frac{d}{d z} \frac{\pa L}{\pa u_z} =0\\
\label{el2}
\frac{\pa L}{\pa v} -\frac{d}{d x} \frac{\pa L}{\pa v_x}
-\frac{d}{d y} \frac{\pa L}{\pa v_y}-\frac{d}{d z} \frac{\pa L}{\pa v_z} =0\\
\label{el3}
\frac{\pa L}{\pa w} -\frac{d}{d x} \frac{\pa L}{\pa w_x}
-\frac{d}{d y} \frac{\pa L}{\pa w_y}-\frac{d}{d z} \frac{\pa L}{\pa w_z} =0
\eea
and
\be
\frac{\pa (x+u,y+v,z+w)}{\pa (x,y,z)}- 1 = 0,
\label{ndiv2}
\ee
which is equivalent to (\ref{ndiv}).  Using the special form
 of $W$ from  (\ref{egy}) without a Lagrange multiplier still
leads to a linear PDE.
Adopting $W$ from (\ref{egy}) is not an essential simplification of the
problem, since the condition
(\ref{ndiv2}) is the main nonlinearity.
The principles of the following calculations apply in the same way to
a nonlinear $W$.

\subsection{Circular twist disclination}

A circular twist disclination is
 a nontrivial example of a topological defect which enforces large deformations
 of an elastic continuum.
The twist disclination
 can be realized as follows  (see fig.~\ref{ctd}):

\begin{figure}[h]
\includegraphics[width=100mm]{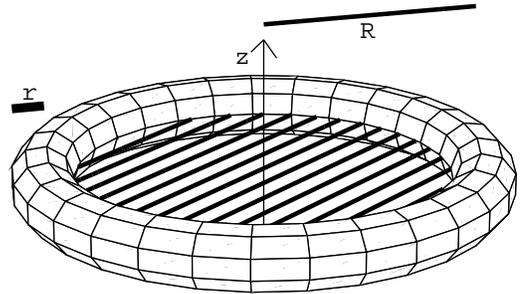}
\caption{Schematic description how to produce a twist disclination
 in an elastic continuum. The solid
torus (size $R$, thickness $r$) is removed.
Then the material is cut along the hatched
surface. After twisting the cut faces by the amount of the Frank angle
$\Om$, the material
is rejoined by glueing.}
\label{ctd}
\end{figure}

Imagine $\R^3$ filled with elastic material and remove a solid torus
centered around the origin and with $z$ as symmetry axis (fig.~\ref{ctd}).
Cut the material along the
surface bounded by the inner circle with radius $R-r$ of the torus
in the $x-y$-plane
(hatched disk in fig.~\ref{ctd}). The cut faces are twisted
with respect to each other by  the Frank angle
$\Om$, and glued together again.

If one shrinks  the removed torus to a singularity line, in the limit $r \to 0$
one obtains a circular twist disclination with radius $R$.
In the subsequent computations,
a sequence of finite values of $r$ is used.

The above description of the twist disclination is equivalent to the one given by
\citeN{Mur:70},
who used however the term 'edge disclination'\fo{To avoid
confusion with screw and edge dislocations, the terms `twist'
and `wedge' are commonly used for disclinations
(\citeNP{Wit:73a}; \citeyearNP{Wit:73b}; \citeNP{Zub:97}). \citeN{Rog:76}
calls the circular twist disclination a `third type of defect' and
Unzicker (\citeyearNP{Unz:96}; \citeyearNP{Unz:00}) a `screw dislocation loop'.
In any case, the twist disclination referred here causes
locally a Volterra distortion of the 5th kind (see \citeNP{Pun:96}).} to
indicate that the Frank vector is perpendicular to the disclination
 line, in analogy to  {\em edge dislocations\/} where the  Burgers vector
 is perpendicular to the defect line.
They investigated analytically  the twist disclination in
linear approximation.

\subsection{Transformation to curvilinear coordinate systems}

The boundary conditions for a specific problem can be
formulated most conveniently in an appropriately chosen
curvilinear coordinate system.  Furthermore, different
scale factors in this coordinate systems allow to pave
the regions of interest - for example regions of large
deformations - more densely with coordinate lines, which is of
great numerical advantage. Frequently, one can use
symmentries of a problem and reduce it to a lower dimension.
The  variety of orthogonal coordinate systems allows
to design an appropriate system for nearly every
problem, even for those with less symmetry. To give some
examples, straight line defects may bemodelled in a cylindrical
system ( e.g. using a log-transformed radial component),
closed line defects with various elliptic coordinate systems.

\begin{figure}
\includegraphics[width=80mm]{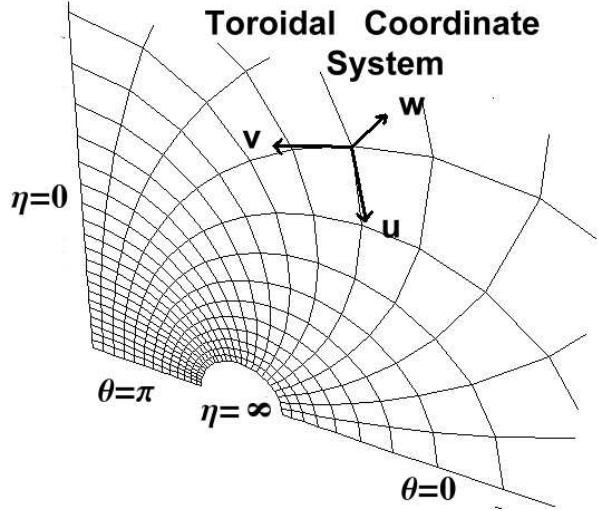}
\caption{Toroidal coordinate system obtained by rotating a bipolar
coordinate system around the $z$-axis ($\ae=0$).
 For symmetry reasons, only
the region $\ae>0$ and $0< \th < \pi$ (at $\ph = 0$) is shown. The value
$\ae= \infty$ corresponds to a focus of the bipolar
system or, after rotating around the $z$-axis, to a circular
singularity line in the toroidal system. $\ae=\th=0$ is the
infinitely far point.}
\label{toroidal}
\end{figure}

For the above described twist disclination, the toroidal system,
as visualized in Fig.~\ref{toroidal},
is a suitable choice. 
To obtain the 3D-toroidal coordinate system from the 2-D
bipolar
system which is actually shown in Fig.~\ref{toroidal},
one must rotate the latter around the vertical axis.

The transformation is given by
\bea
x= \frac{\cos (\p)\,\sinh (\ae)}{-\cos (\th) + \cosh (\ae)}\\
y=  \frac{\sin (\p)\,\sinh (\ae)}{-\cos (\th) + \cosh (\ae)}\\
z=  \frac{\sin (\th)}{-\cos (\th) + \cosh ( \ae)}
\eea

The parameter of the azimuthal rotation is $\phi\in [0,2\pi[$, whereas the polar
angle $\th$ ranges from $0$ to $\pi$. The
hyperbolic coordinate $\ae$ ranges from $-\infty$ (left focus)
to $\infty$ (right focus).

In the region in the vicinity of the circular
singularity the Jacobian determinant is very small.
$\ae = \th = 0$ corresponds to the infinitely far point.
The  components of the displacement vector
in $\ae$-, $\th$-, and $\phi$-direction
are denoted as $u$, $v$ and $w$.

The transformation of differential operators like ${\rm div\,}$ and
$\curl$ in curvilinear
systems is long known (\citeNP{Lov:27}, p.~54) and is
easily done by CAS\footnote{Here the
{\em Mathematica}\copyright \ package `VectorAnalysis'  has been used.}.

To transform the expressions for the energy $W$ and for the
condition (\ref{ndiv2})
in curvilinear coordinates, however, this is not sufficient, since
the displacement vector components $(u,v,w)$ are involved.
 One has to take into account that the basis vectors
in a curvilinear system are subject to change and differentiate
them. This technique is well-known in differential geometry
since it is needed for a curved spacetime in the general theory of
relativity.

The spatial derivatives of the displacement vector
${\bf u } = u^i =(u,v,w)$
are $\frac{\partial {\bf u }}{\partial x_k}$\fo{In contrast to
eqns.~(\ref{el1}-\ref{el3}), $(u,v,w)$ and
$\frac{\partial}{\partial x_k}$ refers to
$\ae$-, $\th$-, and $\phi$-directions in the toroidal system.}.
Taking into account the transformation of the basis vectors,
 the usual derivative has to be
replaced by the `covariant' derivative:
\be
\frac{D }{D x^k} u^i = \frac{d }{d x^k} u^i-\G_{j k}^{\ i}  u^j
\ee
 The
connection $\G_{j k}^{\ i}$ is calculated from
the Jacobian matrix $A_{j}^{\ i}$ ($A_{1}^{\ 1} = 
\frac{\partial \ae}{\partial x}$ etc.) by
\be
\G_{j k}^{\ i} =  B_{k}^{\ i} \frac{\pa}{\pa x^m} A_{j}^{\ m},
\label{CD}
\ee
$B_{i}^{\ j}$ is the inverse of $A_{j}^{\ i}$
(\citeNP{Sch:54}, p.~121ff.).
Eqn.~(\ref{CD}) has to be applied to {\em all} derivatives
occurring in (\ref{el1}-\ref{ndiv2}). This again is done by a
CAS.

The transformation considerably complicates the equations.
For example (\ref{egy}) becomes in toroidal
coordinates
\bea
W =u_{\ae} (\cosh {\ae} -\cos \th  ) +
v_{\th} (\cosh {\ae} -\cos {\th} ) + \\ \nonumber
  u ( \csch \ {\ae} -\cos {\th} \coth \ {\ae} )
 - 2 v \sin {\th}  - u \sinh {\ae}  + \\ \nonumber
  \bigr( w_{\ae}^2 ( \cosh {\ae}-\cos {\th} )^2 +
w_{\th}^2 (-\cos {\th}  + \cosh {\ae} )^2 + \\ \nonumber
     w^2 (\cos {\th} \cosh {\ae} -1)^2 \csch \ {\ae} ^2 +
 w^2 \sin {\th} ^2 + \\ \nonumber
     (v_{\ae} ( \cosh {\ae} -\cos {\th} ) +
 u \sin {\th} )^2 + \\ \nonumber
     (u_{\ae} ( \cosh {\ae} -\cos {\th}) -
 v \sin {\th} )^2 + \\ \nonumber
     (u (\csch \ {\ae} -(\cos {\th} \coth {\ae} ) ) -
 v \sin {\th} )^2 + \\ \nonumber
     (v_{\th} ( \cosh {\ae} -\cos {\th} ) -
 u \sinh {\ae} )^2 + \\ \nonumber
     (u_{\th} (\cosh {\ae} -\cos {\th}) +
 v \sinh {\ae} )^2 \bigl) /2
\eea

The transformation of eqns.~(\ref{el1}-\ref{ndiv2}),
i.e. the full nonlinear PDE to solve, is not written out
here,\fo{Note that
minmization of $W$  in the neo-Hookean case eqn. \ref{egy} still
leads to a linear PDE.  Only the constraint of incompressibility
 (\ref{ndiv}) enforces the substantial nonlinearity of the final PDE.}
they amount to a text file of about 30 $kB$.
The extensive algebraic calculations necessary to transform the
equations are done by a CAS.

\subsection{Boundary conditions}

The circular singularity of the twist disclination
corresponds to $\ae= \infty$.
A toroidal core $\ae > \ae_{max}$ sourrounding
the singularity $\ae = \infty$ is removed from the material.
The chosen values of $\ae_{max}~=~2\ldots 3.625$
correspond to core radii $r~=~0.23\ldots 0.049$.
Since $\ae=\th=0$ is the infinitely far point, only a region
$\ae > \ae_{min}$ and $\th > \th_{min}$ is considered. 

The distance $R$ of the singularity line to the symmetry axis $\ae=0$
(the size of the defect) is set to $1$.

 The cylindrical
symmetry of the problem in the coordinate $\p$ reduces the
3D-problem to two dimensional one. Therefore,  $u$, $v$ and $w$
depend on $\ae$ and $\th$ only. Furthermore, due to mirror symmetry
$\th$ is restricted to the region $0 < \th < \pi$.

The boundary values of $u$, $v$ and $w$ are left
free where ever possible in order to allow relaxing to
the configuration of minimal energy. This is done at
the surface of the torus sourrounding the singularity
at $\ae=\ae_{max}$ (torus shown in fig.~\ref{ctd}), and at
$\ae=0$. The boundary conditions at  $\th=0$
are enforced by the symmetry of the problem and
at $\th=\pi$ by the 'cut-and-glue' condition with the Frank angle $\Om$.

Thus at the disk defined by $\th=\pi$, $v$ vanishes and $u$ and $w$ are
fixed  to a purely circular displacement
corresponding to the Frank angle $\Om$ (see fig.~\ref{toroidal}).
During energy minimization,  the shape
of the removed torus is kept fixed, otherwise the material
could  overlap, which is physically impossible. The boundary values
  $w(\th=\pi, \ae = \ae_{max})=
u(\th=\th_{min}, \ae = \ae_{max})$ implicitly implement this condition.
Only for infinitesimal small values of $\Om$ the $u$ component
would vanish.
In the linear approximation,
$u$, $v$ and $w$ vanish at $\th=0$. At the symmetry
axis $\ae=0$ the condition of incompressibility causes
the vanishing of $u$ and $w$ without an explicit setting to zero.

\section{Numerical methods}

\label{num}
\subsection{Discretization with a \\
difference scheme}

The nonlinear Euler-Lagrange equations generated by the
CAS are discretized
on a twodimensional lattice by evaluating the coefficients
$\ae, \sin \th$ etc. at every lattice point. The displacements
and their derivatives are still maintained in
analytic form.
One obtains 4 variables (displacement
vector ${\bf u }$ and Lagrange multiplier $\l$) at
each of the $n$
grid points. Therefore
$4\,n$ nonlinear equations have to be solved simultaneously.

\subsection{Newton's method in multidimensions}

The solutions of the arising nonlinear system
are obtained by the
Newton-Raphson method of successive approximations
(\citeNP{NR}, chap. 9.7), which is a
multidimensional extension of Newton's root finding
algorithm.
As in one dimension the algorithm starts from an basic
state and uses first derivatives
to iteratively calculate approximations of the solution.

The numerical computation of the $4n$ derivatives
of the Euler-Lagrange equations is however numerically inhibitive.
This problem is solved by
{\em analytically\/} differentiating the Euler-Lagrange equations
at every lattice point with respect
to the displacements $u$,$v$ and $w$ and their
derivatives using a CAS.
Thus the 18 quantities 
 $\l_{\ae}$, $\l_{\th}$,
$u_{\ae}$, $u_{\th}$, $v_{\ae}$,
$v_{\th}$, $w_{\ae}$, $w_{\th}$, $u_{\ae  \th}$, $v_{\ae  \th}$,
$w_{\ae  \th}$, $u_{\th \th}$,
$v_{\ae \ae}$, $w_{\ae  \ae}$, $w_{\th \th}$
entering the Euler-Lagrange equations
 are discretized by
difference molecules (\citeNP{Bro}, p.~767f.).

Only the simplest difference molecules are implemented to avoid
instability of the Newton-Raphson method due to unrealistic smoothness assumptions.
The possible grid dimensions do not allow to assume
smoothness of the solution because
the distorsions in the vicinity of the toroidal core region
usually become very large and  higher order difference operators
contain no next-neighbor coupling.

In addition, the  difference scheme allows for a convenient
 formulation of the boundary conditions.

\subsection{Global convergence}

Although the above implementation is numerically very efficient,
the global convergence of the direct
multidimensional Newton-Raphson method is still not guaranteed.
To obtain the  global convergence
of the algorithm one usually
defines a functional ${\bf V}$ on the solution space
as the square of the l.h. sides of eqns.~\ref{el1}-\ref{ndiv2}.
If the Newton step increases ${\bf V}$
with respect to the previous state,
the algorithm used here `walks back' along the Newton direction
looking for a one-dimensional local minimum of ${\bf V}$.
(cfr. \citeNP{NR}, 9.7). The existence of such a local minimum
is guaranteed, since at the basic state,
${\bf V}$ by definition of the gradient
decreases along the Newton direction.
It turns out that convergence improves if
the `walk back' already is undertaken if ${\bf V}$
decreased slightly, and the Newton step was
`accepted' only if ${\bf V}$ decreases by less than a factor of 10.
A factor 10 is easily reached in the vicinity of the
solution, where the Newton-Raphson algorithm converges
 quadratically with distance.
If the Newton step is accepted, its result
is used as the basic state of the next iteration step
around which the PDE is again linearized.
In contrast to the advice given in \citeN{NR},
it is found here that
it is useful to minimize
${\bf V}$ along the Newton direcetion precisely.
Although this requires more function evaluations,
it performs superior to
taking some premature value for a new linearization.

For the one-dimensional minimization along the Newton direction, the
 golden sectio algorithm (\citeNP{NR}, chap. 10.1) is used.
Where the function is sufficiently flat\fo{And if
the point in the middle was not too
excentric.} (functionals ${\bf V}$ at three
points vary by less than a factor 1.2), the golden
sectio is accompanied by parabolic interpolation
(\citeNP{NR}, chap. 10.2) of the
minimum.
This combination, is found to perform quite well in finding solutions.
Moreover, it is not sensitive to small variations
of the above parameters.

\subsection{Basic states}

\label{ini}

The success of the Newton-Raphson method critically depends
on  the basic state
used for the first linearization. In
the multidimensional case there is little hope
that the algorithm directly converges
to the solution without
the above mentioned global-convergence methods.
But even then
the algorithm ends up in a local minimum
if the first basic state is too far from the solution.
There is for example, no chance to find a solution when starting
with a nonspecialized basic state, e.g. with all
variables set to zero.
Sometimes linear approximations,
e.g. states obtained by assuming small deformations,
are chosen as basic states.
In the present case, the linearized
constraint of incompressibility  $ div \ {\bf u }=0$,
still is too far from
reality to be used for constructing a first basic state.
 Thus,  the incompressibilty condition
is directly implemented by using the symmetries of the
toroidal coordinate system.
In first approximation, it can be assumed that the
displacement takes place along concentric circles
with respect to the
symmetry axis. Only for small displacements this
corresponds to the
tangential displacement of the linearized equation. Thus,
no dilatation at all is allowed in the circular approximation.
For practical reasons,
only the component $w$  of the diplacement is varied.
The components  $u$ and $v$ are adjusted to meet the above requirement.
Thisprocedure requires to solve only a onedimensional linear
PDE for calculating the basic states.

\subsection{Interpolation and extrapolation}

For finer grids and smaller core radii,
 it becomes increasingly difficult
to find solutions by means of the above method.

It turns out that grid refinement by interpolating
a previously found solution
on a coarser grid is by far more efficient than
starting with an independent
new solution of the linearized system.
The  interpolation is performed by a 2D-spline algorithm
(\citeNP{NR}, chap.~3.6) and the interpolated function
 -- evaluated at the refined grid --
is taken as new basic state.
Starting from this state, a new solution usually is
obtained by Newton-Raphson.
However, this refined solution still has the same core
radius as the previous coarse solution.
For extending the grid towards smaller core radii,
that is to larger $\ae_{max}$,
a very efficient method is to attach one lattice line to
the old solution by linear {\em extra\/}polation, and
to use this extended solution again as a new basic state.
All solutions for very small core radii have been obtained in this way.

\subsection{Matrix inversion and function evaluation}

Matrix size increases very rapidly with the number of
grid points. For example, the four functions ($\l, u, v, w$)
on a $30 \times 30$ lattice lead to a $3600 \times 3600$  matrix with
almost $13 \times 10^6$ coefficients, which has to be inverted
at every Newton step. Since the matrix is sparse,
the inversion is more efficiently done by
specialized sparse matrix algorithms.
Here the corresponding packages of
{\em MatLab}\copyright \ are applied over a
 data exchange interface with {\em Mathematica}\copyright.

Since CAS are slow in evaluating trigonometric functions, the
evaluation is done by an external $C$ routine.
This speeds up the computation of the matrix coefficients
by a factor 10 as compared to the CAS.

With this features, satisfactory results are obtained
by running the program on a usual PC.

\section{Results}
\label{res}
\subsection{Linear approximation}

\begin{figure}
\includegraphics[width=80mm]{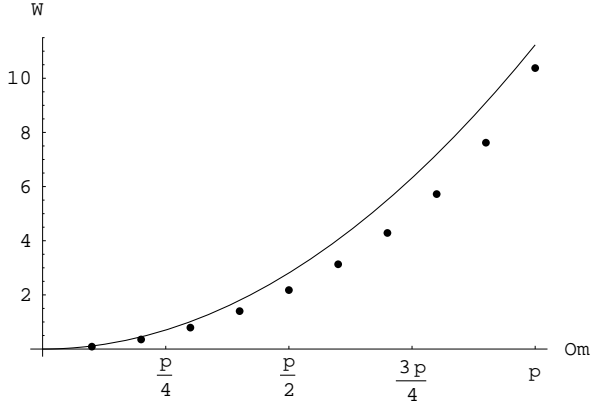}
\caption{Comparision of the numerical results for the linear
approximation (filled circles) with the analytical results (solid line) of
Huang and Mura (1970) in case of a core radius of 0.09 ($\ae_{max}=3$, 
$\ae_{min}=\th_{min}=0.05$, $R=1$, $\mu=3$).
The defect energy increases with the amount
of the Frank angle $\Om$ (ranging from $0$ to $\pi$). For small $\Om$, the results agree.}
\label{eg_om}
\end{figure}

In linear approximation the twist disclination can be
treated analytically. \citeN{Mur:70} have found that the total 
elastic energy is
\bea
\label{egfunc}
W = (\frac{1}{3} \m \ \Om^2 R^3) \{ [2+(1-\frac{r}{R})^2] 
{\rm K \,}(1-\frac{r}{R})-\\ \nonumber
2[1+(1-\frac{r}{R})^2] {\rm E \,}(1-\frac{r}{R}) \},
\eea
where  K and E denote the complete elliptic integral of the first
and second kind. Eqn.~\ref{egfunc} can be approximated by
\be
W = \frac{1}{3} \m \ \Om^2 R^3 (\frac{1}{2} \log(8 R/r)-\frac{4}{3}).
\label{egfuncn}
\ee
This result is used to test  the
algorithm for calculating the basic states (section~\ref{ini}).
Fig.~\ref{eg_om} shows that numerical and analytical results agree
for small $\Om$. The predicted
$\Om^2$-dependence in eqn.~\ref{egfunc} is reproduced by the
linear algorithm.

For larger values of $\Om$, however, the energies
calculated numerically lie below the parabola of
\citeN{Mur:70}.
This illustrates the two aspects of
a linear approximation: both in the analytical
and  in the numerical
treatment shear stresses are assumed to cause pure
shear, but the latter approach lifts the
small-deformation assumption by taking the energy
function (\ref{egy}).

\subsection{Solutions of the complete nonlinear PDE}

\begin{figure}
\includegraphics[width=100mm]{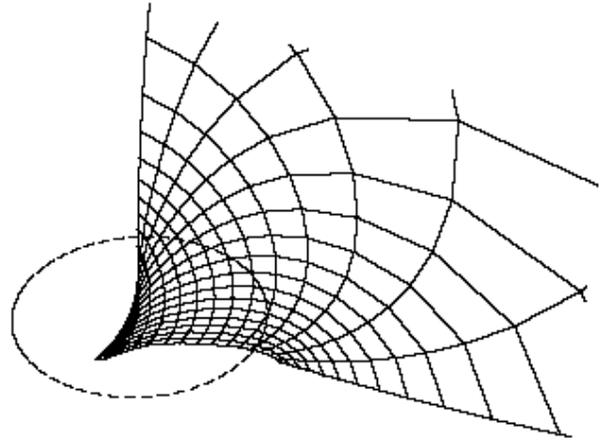}
\caption{Solution of the  nonlinear PDE. The
respective undistorted state is shown in fig.~\ref{toroidal},
with singularity as dottet line.
A very large deformation with a Frank angle $\Om = \pi$
is implemented by twisting the lower boundary $\th= \pi$
by the amount of $\pi/2$. The lower part $-\pi/2 <\th <0$
(not shown here) is twisted in the opposite direction.
The core radius is $r$ about $0.23$, $\ae_{max}=2$, 
$\ae_{min}=\th_{min}= 0.05$.}
\label{frankpi}
\end{figure}

The linear assumption of shear stresses causing shear deformations only is
lifted now.  The solutions of the fully nonlinear equation 
presented here are constraint to the
the case $\Om =\pi$ where very large deformations occur.
Fig.~\ref{frankpi} visualizes these deformations
for a core radius $r=0.23$. The other parameters are $\mu=3,~R=1,
\ae_{min}=\th_{min}= 0.05$. The dottet line indicates the singularity.
Fig.~\ref{toroidal} shows the corresponding undeformed state.

\begin{figure}
\includegraphics[width=70mm]{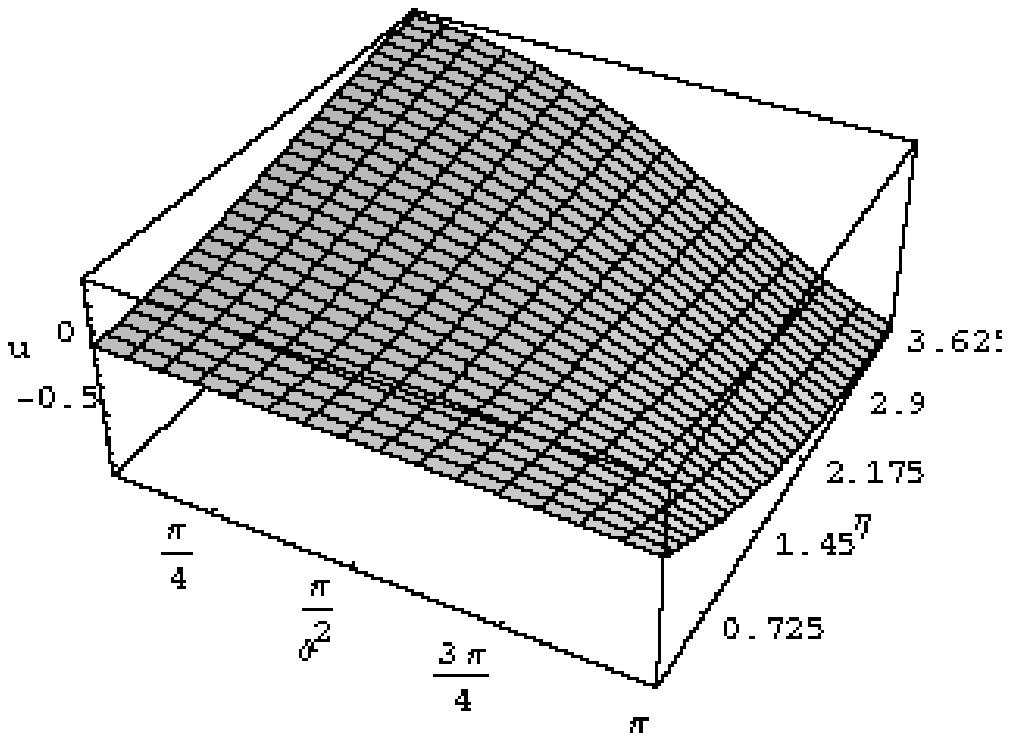}
\includegraphics[width=70mm]{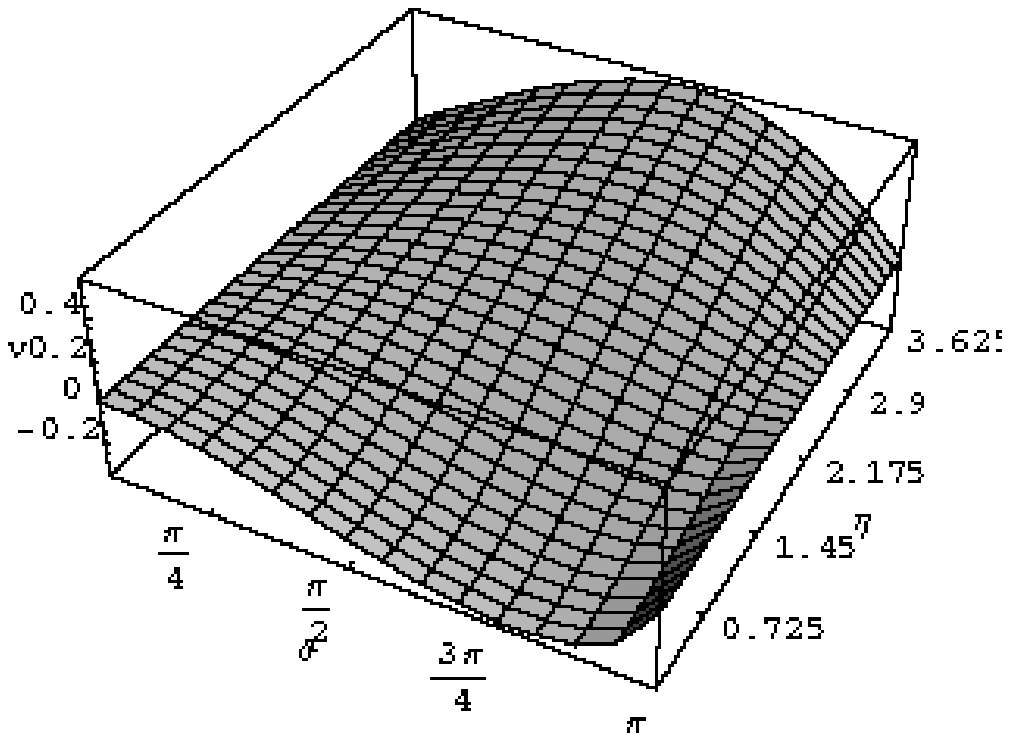}
\includegraphics[width=70mm]{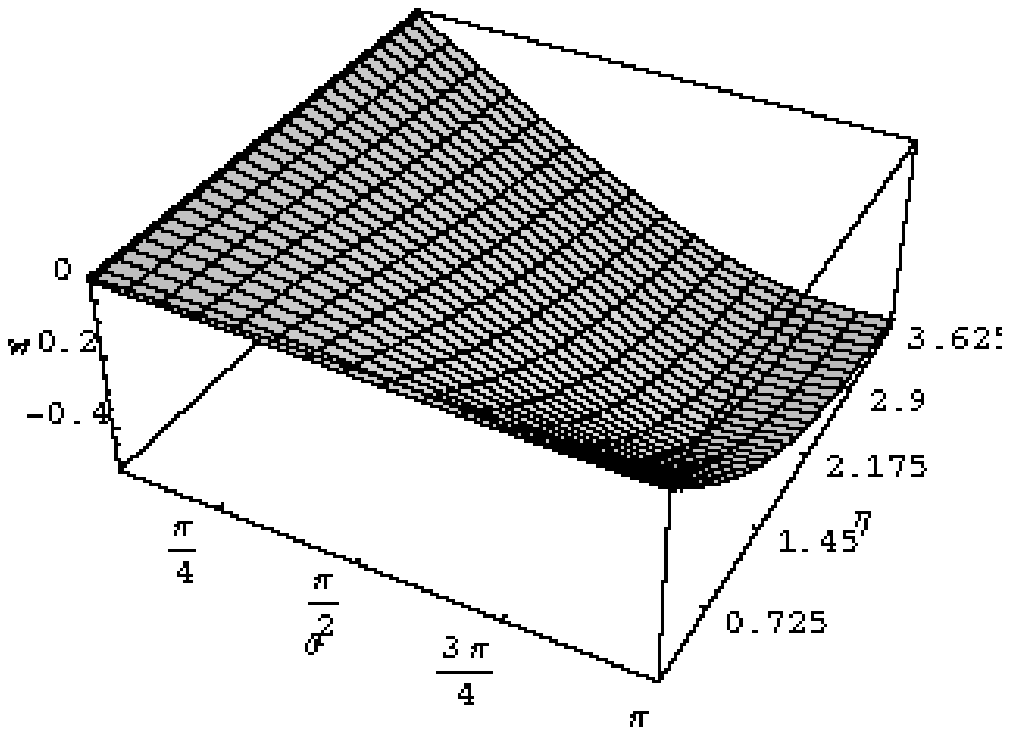}
\caption{
vector components
$u(\ae,\th)$, $v(\ae,\th)$ and $w(\ae,\th)$ for the solution
fig.~\ref{frankpi} (core radius $r=0.23$).}
\label{uvw}
\end{figure}

The components
of the displacement vector $u$, $v$ and $w$,
for the same parameters as in fig.~\ref{frankpi},
are shown in fig.~\ref{uvw}.
Note that the grid shown in fig.~\ref{uvw} is equidistant
in the toroidal coordinates $\ae$ and $\th$, but not in
a cartesian frame, as it is evident from figs.~\ref{toroidal}
and \ref{frankpi}. There, the lattice spacing is much finer in the
region of large deformations.

Even if it demands quite a lot of imagination, the reader
should verify that the displacement components in fig.~\ref{uvw},
attached to fig.~\ref{toroidal}, yield indeed the deformation
fig.~\ref{frankpi}.

\begin{figure}
\includegraphics[width=70mm]{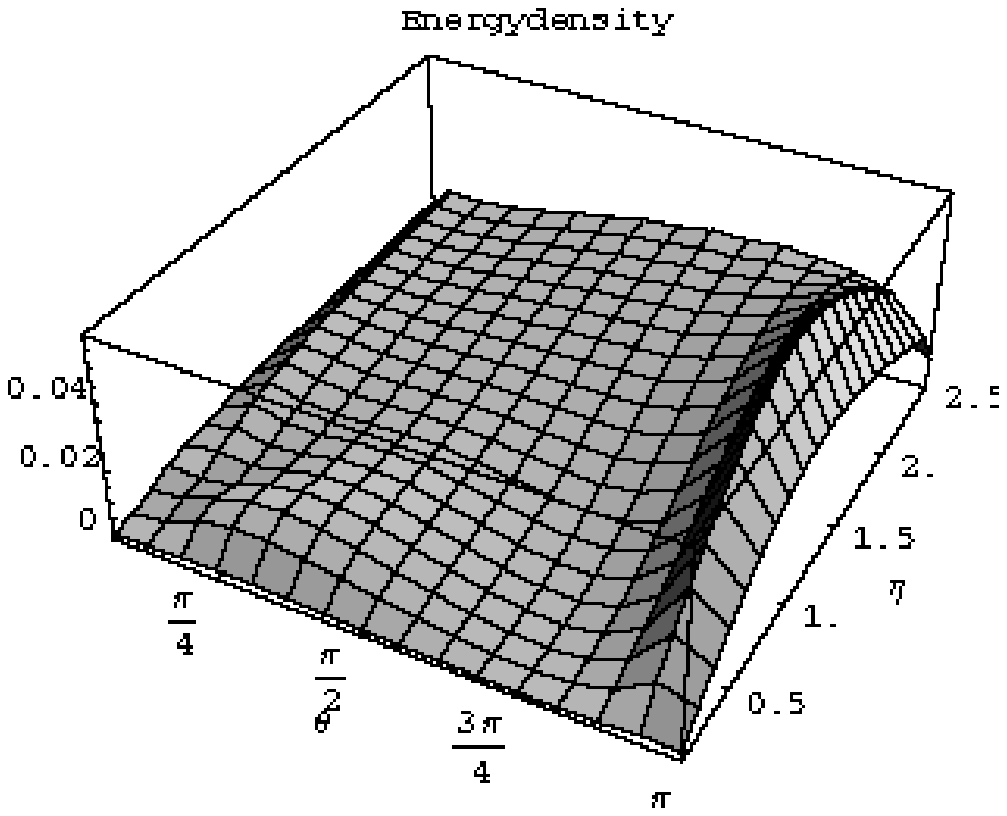}
\caption{Distribution of the elastic energy for an circular twist
disclination with $\ae_{max} = 2.5$ (equivalent to $r=0.145$),
and $\ae_{min}=\th_{min}= 0.05$.}
\label{egden}
\end{figure}

In fig.~\ref{egden} ($r=0.145$) the distribution of the elastic energy
plotted. At each grid point the energy density, weighted by
 the corresponding toroidal  volume
element, is shown.
Most of the energy is stored in the region nearby the cut surface,
where its density is greater than in the vicinity of the core region $\ae=\ae_{max}$.
This effect is even more pronounced for smaller
core radiii $r$.

\subsection{Elastic energy}

\begin{figure}
\includegraphics[width=80mm]{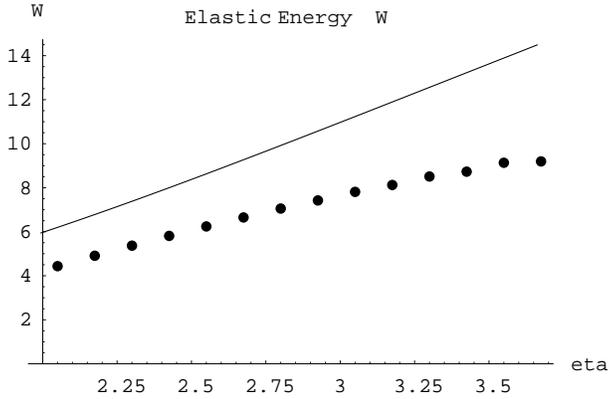}
\caption{Dependence of the total elastic energy
of circular edge disclinations for $\ae_{max}=2.0$
($r = 0.23$) to $\ae_{max}=3.625$ ($r = 0.049$). The
filled circles are obtained from the nonlinear modelling, the
solid line shows the analytic result of Huang and Mura (1970).
The cutoff of the infinitely far point was $\ae_{min}=
\th_{min}=0.05$, which corresponds to a distance of about
20 from the origin. There were 16 lattice points in 
$\th$-direction and a range from 17 to 30 for $\ae$.}
\label{eg_ae}
\end{figure}

 Fig.~\ref{eg_ae} shows the
total elastic energy of a circular twist disclination
in function of $\ae_{max}$ which parametrizes the core radius.

Using the global minimization techniques discussed above,
solutions over a wide
range from $\ae_{max}=2.0$ (which corresponds to
$r = 0.23$) to $\ae_{max}=3.625$ ( or $r = 0.049$)
are found.
Fig.~\ref{eg_ae} shows the dependence of the energy (filled circles)
from $\ae$.\fo{$\ae$ stands for $\ae_{max}+\ae_{min}$ here.}
It is considerably reduced in comparison to
the linear approximation of \citeN{Mur:70} (solid line).
Although from numerical results like those of 
fig.~\ref{eg_ae}
it cannot be deduced that the energy approaches a 
finite value in the limit $r \rightarrow 0$, fig.\ref{egden}
suggests the energy density decreases sufficiently for convergence.

The total elastic energies shown in fig.~\ref{eg_ae}
are calculated from a lattice of 16 points in $\th$-
direction, whereas the number of points in $\ae$-direction
ranged from 17 to 30. The outer limit of the modelled region is defined by
$\ae_{min}=\th_{min}=0.05$. Other solutions
with a different number of grid points, and smaller values
for  $\ae_{min} $ and $ \th_{min}$ (0.01, 0.02) showed 
only a very slight dependence on these parameters.

\subsection{Poynting effect}

The solutions of the nonlinear case show a characteristic
phenomenon which only occurs in the elasticity theory of finite
deformations, the so-called Poynting effect.

The Poynting effect in its classic form is (\citeNP {Tru:65},
p.~193):
`When an incompressible cylinder, free on its outer surface, is
twisted, it experiences an {\em elongation\/} ultimately proportional
to the square of the twist.'

\begin{figure}
\includegraphics[width=70mm]{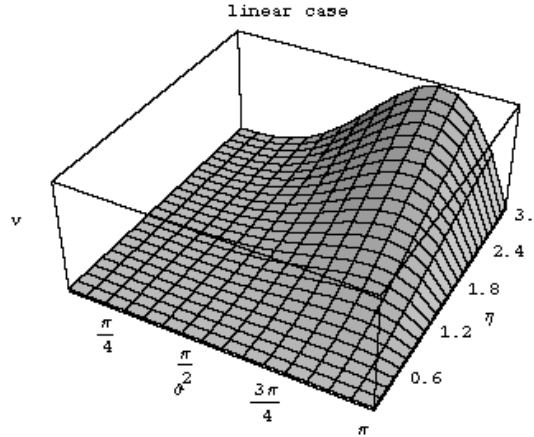}
\caption{Example of a small core radius $r= 0.09$.
Vector component $v$ in $\th$-direction for
a solution of the linearized equations.}
\label{vl}
\end{figure}

\begin{figure}
\includegraphics[width=70mm]{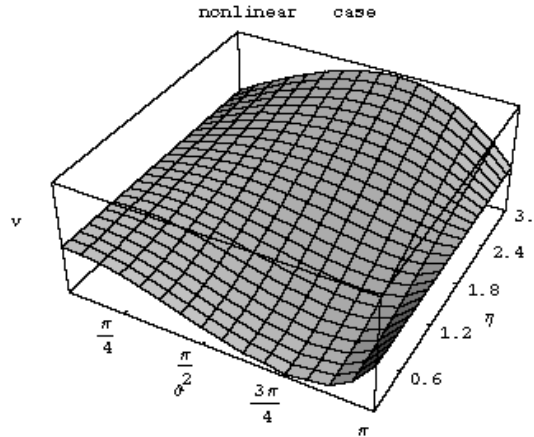}
\caption{As fig.\ref{vl}, but for
the nonlinear case. In the region nearby $\ae=0$, i.e.
in the vicinity of the symmetry axis, $v$ has negative values.
that means it  is pointing upwards. The material is stretched
and the cut surface contracts.}
\label{vn}
\end{figure}                         

To illustrate the difference between the linear approximation
 and the nonlinear solution,
 fig.~\ref{vl} and \ref{vn} show the vector
component $v$ in $\th$-direction for $r=0.09$.
 In the region close to $\ae=0$, i.e.
in the vicinity of the symmetry axis, $v$ has negative values
and the displacement vector is pointing
upwards in the region where $z>0$. (cfr. fig.~\ref{toroidal}).
By mirror symmetry it  points downwards in the region of $z<0$
($-\pi < \th < 0 $). As a consequence the material is stretched and
incompressibility forces  the central region close to the cut 
surface to contract.

The elongation indicated for $z>0$ is observed in the nonlinear case only.
It is a displacement normal to the plane on which the stress is induced
(the cut surface). 
In the linear case from the assumption of pure shear
caused $v=0$ at the symmetry axis (fig.\ref{vl}).

\begin{figure}
\includegraphics[width=80mm]{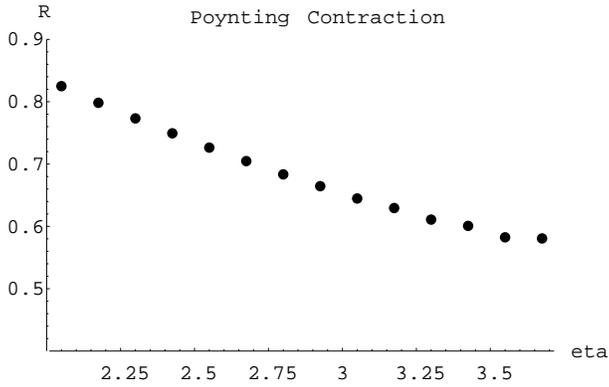}
\caption{Poynting contraction as fraction of the original
size of the defect. The contractions seems to reach a lower bound
at .58, at a core radius of 0.049, which corresponds
to $\ae_{max}=3.625$.}
\label{po_ae}
\end{figure}

The Poynting
contraction apparently approaches a positive
 lower limit for very small core radii $r$. The results shown in 
fig.~\ref{po_ae} suggest that the
contraction does not fall below $58 \%$
of the original radius.

\subsection{Problems}

For both $\ae_{max}<2.0$ and $\ae_{max}>3.75$ it becomes increasingly
difficult to find solutions of the nonlinear equations.
In the first case the difficulty apears to be related to the 
imposed boundary conditions which fix the torus diameter $r$ 
(i.e. the core radius).
In the second case the Poynting contraction
leads to a net change in the torus volume, 
which apparently leads to numerical
problems.

On the other hand, for small core radii, the program 
finds solutions which satisfy the numerical
convergence criterium, but show a physically unreasonable
zigzag- pattern in the displacements.
Possibly the occur in huge deformations
(a volume element of the size of the core radius becomes
stretched to the length of 
the singularity line) become numerically untractable
at least with the present method (`relatively' simple
coordinate system).
However, there still may be a deeper
physical reason for this instability.

A simpler analogue to the circular edge disclination
in case of large deformations is
a twisted rubber band.
Everyday experience shows that while
the Poynting effect only slightly elongates the band,
 the band spontaneously looses
its axial symmetry and coils up as soon as the twist surpasses 
a critical value.

In case of the edge disclination, the
present model -- which essentially depends on 
rotational and mirror symmetry
is not able to correctly model such a transition.
It is therefore not astonishing that meaningful
 solutions are found up to
a maximal $\ae_{max}$ only.

\section{Conclusions}

By combining analytical and numerical approaches, 
a general method for obtaining rigourous solutions 
of nonlinear boundary value problems with nontrivial
geometries has been developed. 

With this new method, for the first time the fully nonlinear
problem of a circular twist disclination in 
an hyperelastic imcompressible material could be treated.

While the extensive calculations necessary for formulating 
Euler-Lagrange-equations in a curvilinear coordinate system
were done by the computer algebra system,  the solution
of the discretized equations was obtained by the
Newton-Raphson method with additional global convergence
features. 
The results give a nonlinear correction to the total elastic energy
obtained in analytical treatments \cite{Mur:70}.
Moreover, the method allowed for the first time to model
the contraction of the singularity line  and to observe
the Poynting effect, which is characteristic for
for large deformations in finite elasticity.
An application of the outlined method to
to other nonlinear PDE's with boundary 
value problems in several dimensions seems possible.

\paragraph{Acknowledgement.}
We are grateful to Dr. Markus Lazar for hints to the literature.

\end{document}